\documentclass[aps,twocolumn,amsmath,amssymb,preprintnumbers,showpacs,superscriptaddress]{revtex4}
\usepackage{amssymb}
\usepackage{dcolumn}
\usepackage{bm}
\usepackage{graphicx}
\usepackage{booktabs}
\usepackage{multirow}

\begin{document}

\newcommand*{\PKU}{School of Physics and State Key Laboratory of Nuclear Physics and
Technology, Peking University, Beijing 100871,
China}\affiliation{\PKU}
\newcommand*{\SDU}{Department of Physics, Shandong University, Jinan, Shandong 250100, China}\affiliation{\SDU}
\newcommand*{\CHEP}{Center for High Energy Physics, Peking University, Beijing 100871, China}\affiliation{\CHEP}

\title{Quark-lepton complementarity revisited}

\author{Xinyi Zhang}\affiliation{\PKU}
\author{Ya-juan Zheng}\affiliation{\PKU}\affiliation{\SDU}
\author{Bo-Qiang Ma}\email{mabq@pku.edu.cn}\affiliation{\PKU}\affiliation{\CHEP}

\begin{abstract}
We reexamine the quark-lepton complementarity (QLC) in nine
angle-phase parametrizations with the latest result of a large
lepton mixing angle $\vartheta_{13}$ from the T2K, MINOS and Double-Chooz experiments. We find that there are still two QLC relations
satisfied in P1, P4 and P6 parametrizations, whereas only one QLC
relation holds in P2, P3, P5 and P9 parametrizations separately. We
also work out the corresponding reparametrization-invariant forms of
the QLC relations and check the resulting expressions with the
experimental data. The results can be viewed as a check of the
validity of the QLC relations, as well as a new perspective into the
issue of seeking for the connection between quarks and leptons.
\end{abstract}


\maketitle



The mixing of fermions remains mysterious in the flavor physics. In the standard model of the particle physics, the mixing is described by mixing matrices which show up in the charged current
interaction. The interaction is described by the following Lagrangian,
\begin{eqnarray}
L&=&-\frac{g}{\sqrt{2}}U^\dagger_L \gamma^\mu V_{\rm CKM} D_L W^+_\mu
\nonumber \\
&&-\frac{g}{\sqrt{2}}E^\dagger_L \gamma^\mu U_{\rm PMNS}N_LW^-_\mu + h.c.,\label{lagrangian}
\end{eqnarray}
where
\begin{eqnarray}
&U_L=(u_L,c_L,t_L)^T; \ \
&D_L=(d_L,s_L,b_L)^T;\nonumber\\
&E_L=(e_L,\mu_L,\tau_L)^T; \ \ &N_L=(\nu_1,\nu_2,\nu_3)^T.\nonumber
\end{eqnarray}
 In Eq.(\ref{lagrangian}), $V_{\rm
CKM}$, namely the Cabibbo-Kobayashi-Maskawa~(CKM) matrix~\cite{CKM},
is the mixing matrix describing the mixing between different
generations of quarks. Correspondingly, $U_{\rm PMNS}$, the
Pontecorvo-Maki-Nakagawa-Sakata (PMNS) matrix~\cite{PMNS}, describes
the misalignment of the flavor eigenstates with the mass eigenstates
of leptons.

This similarity, here we refer to the mixing between generations, combined with the pursuit for unification or symmetry has
motivated speculations on connections between quarks and
leptons~\cite{connect}. Of all the attempts, the quark-lepton
complementarity (QLC)~\cite{smirnov,raidal} has caught much attention for it provides a tempting way to link quarks and leptons. Both its
theoretical base~\cite{raidal} and phenomenological
implications~\cite{phenomenology} have been discussed in the
literatures.

The original QLC corresponds to two numerical
relations between the mixing angles of the CKM matrix and the PMNS
matrix, namely,
\begin{eqnarray}
\theta_{12}+\vartheta_{12}=45^\circ,\quad
\theta_{23}+\vartheta_{23}=45^\circ,
\end{eqnarray}
where $\theta_{ij}$ and $\vartheta_{ij}$ denote the mixing angles of
the CKM matrix and the PMNS matrix separately in the standard
parametrization~\cite{CK}. The mixing matrix is a unitary matrix
that can be parametrized by three Euler angles and a CP violating
phase. Such kind of parametrization can be referred to as
angle-phase parametrization. For neutrinos of the Majorana type, two
additional CP-violating phases are needed. Since the Majorana CP-violating phases do not manifest themselves in the oscillation, we
discuss the Dirac neutrinos only. The three Euler angles correspond
to three rotations in complex planes. There is a freedom in
arranging the orders of three rotations and different orders result
in different angle-phase parametrizations. There are nine
angle-phase parametrizaions that are structurally
different~\cite{9/12}. Notice that the QLC relations are
parametrization-dependent~\cite{Jarlskog,Zheng10}. Though different
parametrizations are equivalent to each other mathematically, there
are differences in revealing some phenomenological relations, e.g.,
the QLC relations. Such differences make them of different
significance in analysis and model building.

As a result, it is meaningful to reexamine the QLC relations in nine angle-phase parametrizations
especially when several experiments observe a relatively large
lepton mixing angle $\vartheta_{13}$~\cite{expt}. Such a result
deviates from the previous thought of a quasivanishing one. Another
motivation is that there is still a chance that the QLC relations are
purely accidental, so evaluating the relations is the first step to
take before going further along such a direction.

\break

The starting point is the moduli of mixing matrices, which is
\begin{widetext}
\begin{eqnarray}
|V_{\rm CKM}|= \left(
  \begin{array}{ccc}
    0.97428\pm0.00015   & 0.2253\pm0.0007           & 0.00347^{+0.00016}_{-0.00012}\\
    0.2252\pm0.0007     & 0.97345^{+0.00015}_{-0.00016}  & 0.0410^{+0.0011}_{-0.0007}\\
    0.00862^{+0.00026}_{-0.00020}& 0.0403^{+0.0011}_{-0.0007}& 0.999152^{+0.000030}_{-0.000045}
  \end{array} \right)\;\label{ckmdata}
\end{eqnarray}
for quarks~\cite{pdg} and
\begin{eqnarray}
|U_{\rm PMNS}|= \left(
  \begin{array}{ccc}
   0.824^{+0.011(+0.032)}_{-0.010(-0.032)}& 0.547^{+0.016(+0.047)}_{-0.014(-0.044)}
   &  0.145^{+0.022(+0.065)}_{-0.031(-0.113)}    \\
   0.500^{+0.027(+0.076)}_{-0.021(-0.071)}& 0.582^{+0.050(+0.139)}_{-0.023(-0.069)}
   &  0.641^{+0.061(+0.168)}_{-0.023(-0.063)}     \\
   0.267^{+0.044(+0.123)}_{-0.027(-0.088)}& 0.601^{+0.048(+0.133)}_{-0.022(-0.069)}
   &  0.754^{+0.052(+0.143)}_{-0.020(-0.054)}
  \end{array} \right)
\end{eqnarray}
\end{widetext}
for leptons, which is the latest global fitting results of pre-DAYA-BAY experiments~($1\sigma~(3\sigma)$)~\cite{repara,global}. There are novel measurements of the neutrino
mixing angle $\vartheta_{13}$ by the Daya-Bay
Collaboration~\cite{Daya-Bay} and the RENO Collaboration~\cite{Ahn:2012nd} recently. The Daya-Bay
Collaboration releases $\sin^22\vartheta_{13}=0.092\pm0.016(\mathrm{stat}) \pm 0.005
(\mathrm{syst})$ of a significance of 5.2~$\sigma$, and the
corresponding angle is
$\vartheta_{13}=(8.828\pm0.793(\mathrm{stat})\pm0.248(\mathrm{syst}))^\circ$.  The RENO Collaboration releases $\sin^22\vartheta_{13}=0.113\pm0.013(\mathrm{stat}) \pm 0.019
(\mathrm{syst})$ of a significance of 4.9~$\sigma$, and the
corresponding angle is
$\vartheta_{13}=(9.821\pm0.588(\mathrm{stat})\pm0.860(\mathrm{syst}))^\circ$.
The value of $\vartheta_{13}$ in our analysis is based on the global
fit in Ref.~\cite{global,repara}, with $\vartheta_{13}=(8.332\pm
1.399(\pm4.396))^\circ$, which corresponds to $\sin^22\vartheta_{13}=0.082\pm0.027(\pm0.084)$. Therefore, our analysis is compatible with
the new data.

We calculate the mixing angles of the nine angle-phase
parametrizations with matrix elements that are independent of the CP-violating phase. For example, from the P1, i.e., the standard parametrization, we have,
\begin{eqnarray}
\sin\theta_{13}=|V_{ub}|,\quad\tan\theta_{12}=\frac{|V_{us}|}{|V_{ud}|},\quad\tan\theta_{23}=\frac{|V_{cb}|}{|V_{tb}|}.
\end{eqnarray}
Thus we get the corresponding values of the mixing angles. The
results are listed in Table~\ref{tab:QLC}.

\begingroup
\squeezetable
\begin{table*}[htbp]
 \caption{\label{tab:QLC}The angle-phase parametrizations and quark-lepton complementarity}
 \begin{ruledtabular}
  \begin{tabular}{ccc}
   \toprule
 Parametrization   & Mixing angles & Quark-lepton complementarity \\
 \hline
 &&\\
 \underline{P1:~$V=R_{23}(\theta_{23})R_{31}(\theta_{13},\phi)R_{12}(\theta_{12})$}& $\theta_{12}\slash\theta_{23}\slash\theta_{13}$ \quad
 $\vartheta_{12}\slash\vartheta_{23}\slash\vartheta_{13}$
 &\\
 \multirow{3}{*}{
  $\left(
   \begin{array}{ccc}
   c_{12}c_{13} & s_{12}c_{13} & s_{13} \\
   -c_{12}s_{23}s_{13}-s_{12}c_{23}e^{-i\phi} &
   -s_{12}s_{23}s_{13}+c_{12}c_{23}e^{-i\phi} &
   s_{23}c_{13} \\
   -c_{12}c_{23}s_{13}+s_{12}s_{23}e^{-i\phi} &
   -s_{12}c_{23}s_{13}-c_{12}s_{23}e^{-i\phi} &
    c_{23}c_{13}\\
   \end{array}
   \right)$}
 &$(13.02\pm0.039)^\circ+(33.58^{+0.849}_{-0.748})^\circ=(46.60^{+0.888}_{-0.787})^\circ$
 &$\theta_{12}+\vartheta_{12}\simeq45^\circ$\footnote{$``\simeq"$ refers to $45^\circ$ being in $2\sigma$ error range.}\\
 &$(2.35^{+0.063}_{-0.040})^\circ+(40.37^{+2.880}_{-1.227})^\circ=(42.72^{+2.943}_{-1.267})^\circ$
 &$\theta_{23}+\vartheta_{23}\simeq45^\circ$\footnote{$``\simeq"$ refers to $45^\circ$ being in $1\sigma$ error range.}\\
 &$(0.20\pm0.009)^\circ+(8.33\pm1.40)^\circ=(8.53\pm1.409)^\circ$
 &\\
 &&\\
 \underline{P2:~$V=R_{12}(\theta_3)R_{23}(\theta_2,\phi)R_{12}^{-1}(\theta_1)$}& $\theta_1\slash\theta_2\slash\theta_3 $\quad
 $\vartheta_1\slash\vartheta_2\slash\vartheta_3$
 &\\
 \multirow{3}{*}{
  $\left(
   \begin{array}{ccc}
   s_1 c_2 s_3 +c_1 c_3 e^{-i\phi}& c_1 c_2 s_3-s_1 c_3  e^{-i\phi} & s_2 s_3    \\
   s_1 c_2 c_3-c_1 s_3 e^{-i\phi} & c_1 c_2 c_3+s_1 s_3 e^{-i\phi} & s_2 c_3    \\
   -s_1 s_2                       & -c_1 s_2                     & c_2        \\
   \end{array}
   \right)$}
 &$(12.07^{+0.477}_{-0.340})^\circ+(23.95^{+3.895}_{-2.287})^\circ=(36.02^{+4.372}_{-2.627})^\circ$
 &\\
 &$(2.36^{+0.042}_{-0.063})^\circ+(41.06^{+4.538}_{-1.745})^\circ=(43.42^{+4.580}_{-1.808})^\circ$
 &$\theta_2+\vartheta_2\simeq45^\circ$\footnotemark[2]\\
 &$(4.84^{+0.257}_{-0.186})^\circ+(12.75^{+2.209}_{-2.674})^\circ=(17.59^{+2.466}_{-2.860})^\circ$
 &\\
 &&\\
 \underline{P3:~$V=R_{23}(\theta_2)R_{12}(\theta_1,\phi)R_{23}^{-1}(\theta_3)$}& $\theta_1\slash\theta_2\slash\theta_3 $\quad
 $\vartheta_1\slash\vartheta_2\slash\vartheta_3$
 &\\
 \multirow{3}{*}{
  $\left(
   \begin{array}{ccc}
   c_1      & s_1 c_3                        & -s_1 s_3   \\
   -s_1 c_2 & c_1 c_2 c_3+s_2 s_3e^{-i\phi}  & -c_1 c_2 s_3+s_2 c_3e^{-i\phi}\\
   s_1 s_2  & -c_1 s_2 c_3+c_2 s_3e^{-i\phi} & c_1 s_2 s_3+c_2 c_3e^{-i\phi} \\
   \end{array}
   \right)$}
 &$(13.02\pm0.038)^\circ+(34.51^{+1.113}_{-1.012})^\circ=(47.53^{+1.151}_{-1.050})^\circ$
 &$\theta_1+\vartheta_1\simeq45^\circ$\footnotemark[1]\\
 &$(2.19^{+0.066}_{-0.051})^\circ+(28.10^{+4.131}_{-2.608})^\circ=(30.29^{+4.197}_{-2.659})^\circ$
 &\\
 &$(0.88^{+0.041}_{-0.031})^\circ+(14.85^{+2.194}_{-3.057})^\circ=(15.73^{+2.235}_{-3.088})^\circ$
 &\\
 &&\\
 \underline{P4:~$V=R_{23}(\theta_2)R_{12}(\theta_1,\phi)R_{31}^{-1}(\theta_3)$}& $\theta_1\slash\theta_2\slash\theta_3 $\quad
 $\vartheta_1\slash\vartheta_2\slash\vartheta_3$
 &\\
 \multirow{3}{*}{
  $\left(
   \begin{array}{ccc}
   c_1 c_3                        & s_1      & -c_1 s_3   \\
   -s_1 c_2 c_3+s_2 s_3 e^{-i\phi}& c_1 c_2  & s_1 c_2 s_3+s_2 c_3 e^{-i\phi}  \\
   s_1 s_2 c_3+c_2 s_3 e^{-i\phi} & -c_1 s_2 & -s_1 s_2 s_3+c_2 c_3 e^{-i\phi} \\
  \end{array}
  \right)$}
 &$(13.02\pm0.041)^\circ+(33.16^{+1.096}_{-0.959})^\circ=(46.18^{+1.137}_{-1.000})^\circ$
 &$\theta_1+\vartheta_1\simeq45^\circ$\footnotemark[1]\\
 &$(2.37^{+0.065}_{-0.051})^\circ+(45.92^{+3.360}_{-1.543})^\circ=(48.29^{+3.425}_{-1.584})^\circ$
 &$\theta_2+\vartheta_2\simeq45^\circ$\footnotemark[1]\\
 &$(0.20^{+0.009}_{-0.007})^\circ+(9.98^{+1.490}_{-2.095})^\circ=(10.18^{+1.499}_{-2.102})^\circ$
 &\\
 &&\\
 \underline{P5:~$V=R_{31}(\theta_3)R_{23}(\theta_2,\phi)R_{12}^{-1}(\theta_1)$}& $\theta_1\slash\theta_2\slash\theta_3 $\quad
 $\vartheta_1\slash\vartheta_2\slash\vartheta_3$
 &\\
 \multirow{3}{*}{
  $\left(
   \begin{array}{ccc}
   -s_1 s_2 s_3+c_1 c_3 e^{-i\phi} & -c_1 s_2 s_3-s_1 c_3 e^{-i\phi}  & c_2 s_3   \\
   s_1 c_2                         & c_1 c_2                          & s_2       \\
   -s_1 s_2 c_3-c_1 s_3 e^{-i\phi} & -c_1 s_2 c_3+s_1 s_3 e^{-i\phi}  & c_2 c_3   \\
   \end{array}
   \right)$}
 &$(13.03\pm0.039)^\circ+(40.67^{+2.875}_{-1.634})^\circ=(53.70^{+2.912}_{-1.673})^\circ$
 &\\
 &$(2.35^{+0.063}_{-0.040})^\circ+(39.87^{+4.556}_{-1.718})^\circ=(42.22^{+4.619}_{-1.758})^\circ$
 &$\theta_2+\vartheta_2\simeq45^\circ$\footnotemark[2]\\
 &$(0.20^{+0.009}_{-0.007})^\circ+(10.89^{+1.772}_{-2.290})^\circ=(11.09^{+1.781}_{-2.297})^\circ$
 &\\
 &&\\
 \underline{P6:~$V=R_{12}(\theta_1)R_{31}(\theta_3,\phi)R_{23}^{-1}(\theta_2)$}& $\theta_1\slash\theta_2\slash\theta_3 $\quad
 $\vartheta_1\slash\vartheta_2\slash\vartheta_3$
 &\\
 \multirow{3}{*}{
  $\left(
   \begin{array}{ccc}
   c_1 c_3   & c_1 s_2 s_3+s_1 c_2 e^{-i\phi}  & c_1 c_2 s_3-s_1 s_2 e^{-i\phi}   \\
   -s_1 c_3  & -s_1 s_2 s_3+c_1 c_2 e^{-i\phi} & -s_1 c_2 s_3-c_1 s_2 e^{-i\phi}  \\
   -s_3    & s_2 c_3                     & c_2 c_3                      \\
   \end{array}
   \right)$}
 &$(13.02\pm0.039)^\circ+(31.25^{+1.414}_{-1.111})^\circ=(44.27^{+1.453}_{-1.150})^\circ$
 &$\theta_1+\vartheta_1\simeq45^\circ$\footnotemark[2]\\
 &$(2.31^{+0.063}_{-0.040})^\circ+(38.56^{+2.948}_{-1.263})^\circ=(40.87^{+3.011}_{-1.303})^\circ$
 &$\theta_2+\vartheta_2\simeq45^\circ$\footnotemark[1]\\
 &$(0.49^{+0.015}_{-0.011})^\circ+(15.49^{+2.617}_{-1.606})^\circ=(15.98^{+2.632}_{-1.617})^\circ$
 &\\
 &&\\
 \underline{P7:~$V=R_{31}(\theta_3)R_{12}(\theta_1,\phi)R_{31}^{-1}(\theta_2)$}& $\theta_1\slash\theta_2\slash\theta_3 $\quad
 $\vartheta_1\slash\vartheta_2\slash\vartheta_3$
 &\\
 \multirow{3}{*}{
   $\left(
    \begin{array}{ccc}
    c_1 c_3 c_2+s_3 s_2 e^{-i\phi} & s_1 c_3  & -c_1 c_3 s_2+s_3 c_2 e^{-i\phi}  \\
    -s_1 c_2                       & c_1      & s_1 s_2                          \\
    -c_1 s_3 c_2+c_3 s_2 e^{-i\phi}& -s_1 s_3 & c_1 s_3 s_2+c_3 c_2 e^{-i\phi}   \\
    \end{array}
    \right)$}
 &$(13.23^{+0.038}_{-0.040})^\circ+(54.41^{+3.524}_{-1.621})^\circ=(67.64^{+3.562}_{-1.661})^\circ$
 &\\
 &$(10.32^{+0.273}_{-0.175})^\circ+(52.04^{+3.042}_{-1.536})^\circ=(62.36^{+3.315}_{-1.711})^\circ$
 &\\
 &$(10.14^{+0.273}_{-0.175})^\circ+(47.69^{+2.427}_{-1.275})^\circ=(57.83^{+2.700}_{-1.450})^\circ$
 &\\
 &&\\
 \underline{P8:~$V=R_{12}(\theta_1)R_{23}(\theta_2,\phi)R_{31}(\theta_3)$}& $\theta_1\slash\theta_2\slash\theta_3 $\quad
 $\vartheta_1\slash\vartheta_2\slash\vartheta_3$
 &\\
 \multirow{3}{*}{
   $\left(
    \begin{array}{ccc}
    -s_1 s_2 s_3+c_1 c_3 e^{-i\phi}& s_1 c_2  & s_1 s_2 c_3+c_1 s_3 e^{-i\phi}   \\
    -c_1 s_2 s_3-s_1 c_3 e^{-i\phi}& c_1 c_2  & c_1 s_2 c_3-s_1 s_3 e^{-i\phi}   \\
    -c_2 s_3                       & -s_2     & c_2 c_3                      \\
    \end{array}
    \right)$}
 &$(13.02\pm0.039)^\circ+(43.22^{+2.596}_{-1.347})^\circ=(56.25^{+2.635}_{-1.386})^\circ$
 &\\
 &$(2.31^{+0.063}_{-0.040})^\circ+(36.94^{+3.443}_{-1.578})^\circ=(39.25^{+3.506}_{-1.618})^\circ$
 &\\
 &$(0.49^{+0.015}_{-0.011})^\circ+(19.50^{+3.222}_{-1.886})^\circ=(19.99^{+3.237}_{-1.897})^\circ$
 &\\
 &&\\
 \underline{P9:~$V=R_{31}(\theta_3)R_{12}(\theta_1,\phi)R_{23}(\theta_2)$}& $\theta_1\slash\theta_2\slash\theta_3 $\quad
 $\vartheta_1\slash\vartheta_2\slash\vartheta_3$
 &\\
 \multirow{3}{*}{
   $\left(
    \begin{array}{ccc}
    c_1 c_3  & s_1 c_2 c_3-s_2 s_3 e^{-i\phi}& s_1 s_2 c_3+c_2 s_3 e^{-i\phi} \\
    -s_1     & c_1 c_2                       & c_1 s_2                        \\
    -c_1 s_3 &-s_1 c_2 s_3-s_2 c_3 e^{-i\phi}& -s_1 s_2 s_3+c_2 c_3 e^{-i\phi}\\
    \end{array}
    \right)$}
 &$(13.01\pm0.041)^\circ+(30.00^{+1.787}_{-1.390})^\circ=(43.01^{+1.828}_{-1.431})^\circ$
 &$\theta_1+\vartheta_1\simeq45^\circ$\footnotemark[1]\\
 &$(2.41^{+0.065}_{-0.041})^\circ+(47.76^{+3.658}_{-1.523})^\circ=(50.17^{+3.723}_{-1.564})^\circ$
 &\\
 &$(0.51^{+0.015}_{-0.012})^\circ+(17.95^{+2.779}_{-1.712})^\circ=(18.46^{+2.794}_{-1.724})^\circ$
 &\\
      \bottomrule
\end{tabular}
\end{ruledtabular}
\end{table*}
\endgroup
From Table~\ref{tab:QLC}, we can see that the QLC is satisfied approximately in seven of the
nine parametrizations. By ``satisfying", we mean $45^\circ$ being in
$2 \sigma$ error range. Of the seven parametrizations that have at
least one QLC relation, the QLC relations for two pairs of mixing
angles hold in P1, P4 and P6 parametrizations, of which P1
corresponds to the standard parametrization~\cite{CK}.

Combined with earlier work on the self-complementarity of the lepton
mixing angles~\cite{Zhang:2012xu}, we find that a parametrization
which has the self-complementarity also has at least one of the QLC
relations. To be explicit, the P1, P3, P4, P6 and P9
parametrizations hold both the lepton self-complementarity and the
quark-lepton complementarity. We see that except for the
well-examined P1 (standard) and P3 (Kobayashi-Maskawa)
parametrizations, P4, P6 and P9 parametrizations stand out as they
have the advantage of satisfying both complementarities. The
self-complementarity relation may result in new mixing patterns. The
PMNS matrix can be expanded around such patterns in orders of the
Wolfenstein parameter $\lambda$ by using the QLC relation. One such example is given in Ref.~\cite{Zheng:2011uz}.

Table~\ref{tab:QLC} can be viewed as an update of the work Zheng did
in Ref.~\cite{Zheng10}. Compared with the results in
Ref.~\cite{Zheng10}, we find that the situation has been changed a
lot. Only one QLC relation holds in more parametrizations whereas
the original form of the QLC, namely two complementarity relations
is satisfied in P1, P4 and P6 parametrizations. Additionally, the
parametrizations that the QLC relations hold do not have a simple
form in their (1,3) entries in common.

As the matrix elements are more relevant to physical observables, we
seek relations of the matrix elements which are
reparamentrization invariant, as Ref.\cite{repara} did for the
standard parametrization only.

By assuming that the QLC relations are exact, we translate the
relations into the form in terms of the matrix elements. For
example, in P2 parametrization we have
\begin{eqnarray}
\theta_2+\vartheta_2\simeq45^\circ,
\end{eqnarray}
since
\begin{eqnarray}
\cos\theta_2=|V_{tb}|,\quad\cos\vartheta_2=|U_{\tau3}|,
\end{eqnarray}
by substituting the trigonometric function with the modulus of the
matrix elements, we have
\begin{eqnarray}
|U_{\tau3}|=\frac{1}{\sqrt{2}}(\sqrt{1-|V_{tb}|^2}+|V_{tb}|).
\end{eqnarray}
We list the results in Table~\ref{tab:moduli}.

\begin{table*}[htbp]
 \caption{\label{tab:moduli}The reparametrization-invariant forms of the QLC relations and their verification}
 \begin{ruledtabular}
 \begin{tabular}{ccc}
   \toprule
      Parametrization& Reparametrization-invariant form& Verification\\
   \hline
   &&\\
   P1 & $\frac{|U_{e2}|}{|U_{e1}|}=\frac{|V_{ud}|-|V_{us}|}{|V_{ud}|+|V_{us}|}$
   &$0.664^{+0.021}_{-0.019},\quad0.62437\pm0.00095$\\
   &&\\
   P1 &$\frac{|U_{\mu3}|}{|U_{\tau3}|}=\frac{|V_{tb}|-|V_{cb}|}{|V_{tb}|+|V_{cb}|}$
   &$0.850^{+0.100}_{-0.038},\quad0.92117^{+0.00209}_{-0.00138}$\\
   &&\\
   P2 & $|U_{\tau3}|=\frac{1}{\sqrt{2}}(\sqrt{1-|V_{tb}|^2}+|V_{tb}|)$
   &$0.754^{+0.052}_{-0.020},\quad0.73562^{+0.00049}_{-0.00074}$\\
   &&\\
   P3 & $|U_{e1}|=\frac{1}{\sqrt{2}}(\sqrt{1-|V_{ud}|^2}+|V_{ud}|)$
   &$0.824^{+0.011}_{-0.010},\quad0.84826\pm0.00035$\\
   &&\\
   P4 & $|U_{e2}|=\frac{1}{\sqrt{2}}(\sqrt{1-|V_{us}|^2}-|V_{us}|)$
   &$0.547^{+0.016}_{-0.014},\quad0.52962\pm0.00061$\\
   &&\\
   P4 &$\frac{|U_{\tau2}|}{|U_{\mu2}|}=\frac{|V_{cs}|-|V_{ts}|}{|V_{cs}|+|V_{ts}|}$
   &$1.033^{+0.121}_{-0.056},\quad0.92049^{+0.00208}_{-0.00133}$\\
   &&\\
   P5 & $|U_{\mu3}|=\frac{1}{\sqrt{2}}(\sqrt{1-|V_{cb}|^2}-|V_{cb}|)$
   &$0.641^{+0.061}_{-0.023},\quad0.67752^{+0.00081}_{-0.00052}$\\
   &&\\
   P6 & $\frac{|U_{\mu1}|}{|U_{e1}|}=\frac{|V_{ud}|-|V_{cd}|}{|V_{ud}|+|V_{cd}|}$
   &$0.607^{+0.034}_{-0.027},\quad0.62450\pm0.00095$\\
   &&\\
   P6 &$\frac{|U_{\tau2}|}{|U_{\tau3}|}=\frac{|V_{tb}|-|V_{ts}|}{|V_{tb}|+|V_{ts}|}$
   &$0.797^{+0.084}_{-0.036},\quad0.92246^{+0.00203}_{-0.00129}$\\
   &&\\
   P9 & $|U_{\mu1}|=\frac{1}{\sqrt{2}}(\sqrt{1-|V_{cd}|^2}-|V_{cd}|)$
   &$0.500^{+0.027}_{-0.021},\quad0.52970\pm0.00061$\\
   &&\\
 \bottomrule
  \end{tabular}
 \end{ruledtabular}
\end{table*}
From Table~\ref{tab:moduli}, we see that the relations can be
generally divided into two kinds. One is a one-to-one relation as in
P2, P3, P4, P5 and P9 parametrizations, while the other kind is the
two-to-two relations as in P1, P4 and P6 parametrizations. The same
expressions for P1 have been pointed out in Ref.~\cite{repara}.
Notice that for central values, there are two cases with larger
deviations, i.e., the second relation in P4 and P6. Such deviations
are a natural result of a relatively large deviation from the exact
numerical QLC relation, which can be seen in Table~\ref{tab:QLC}.

The moduli of the CKM matrix elements, i.e., $|V_{ij}|$ are measured
to a high precision by various processes. Using the QLC relations in
the reparametrization-invariant forms in Table~\ref{tab:moduli}, we
can get information on the relatively not-so-well-determined PMNS
matrix elements $|U_{ij}|$. All of the relations in the second
column of Table~\ref{tab:moduli} are candidates of
reparametrization-invariant QLC relations and their validity could
be tested by future experiments. As there is still chance that the
QLC relations are accidental, the reparametrization-invariant forms
of the QLC relations can also be used as the test for the
validity of the QLC relations with the advantage of being more
directly related to the observables.

As the parametrizations are independent of each other, the
corresponding relations deduced from different parametrizations do
not have to be satisfied simultaneously. In fact, when assuming that
they are satisfied at the same time, unexpected results may emerge.
For example, if we assume that the one-to-one relations of
$|U_{\mu3}|$ and $|U_{\tau3}|$ hold at the same time, by a usage of
the unitarity relation, we will find that $|U_{e3}|$ is purely
imaginary, which contradicts the data.

To sum up, we reexamine the QLC relations
in nine angle-phase parametrizations, and work out the corresponding
relations of the reparametrization-invariant form. We find that the
new experimental data have changed the situation whether a given parametrization has the quark-lepton
complementarity or not and whether a given parametrization has one relation or two.
The reparametrization-invariant form of the QLC relations may suggest some
general connections between quarks and leptons mixing. We look forward to the experimental tests of these relations.

This work is partially
supported by National Natural Science Foundation of China (Grants
No.~11021092, No.~10975003, No.~11035003, and No.~11120101004) and
by the Research Fund for the Doctoral Program of Higher Education
(China).


\begin{thebibliography}{99}

\bibitem{CKM}
N.~Cabibbo, Phys.\ Rev.\ Lett.\  {\bf 10},  531 (1963);\\
M.~Kobayashi and T.~Maskawa, Prog.\ Theor.\ Phys.\  {\bf 49}, 652
(1973).

\bibitem{PMNS}
B.~Pontecorvo, Sov.\ Phys.\ JETP {\bf 26},  984 (1968);\\
Z.~Maki, M.~Nakagawa and S.~Sakata, Prog.\ Theor.\ Phys.\  {\bf 28},
870 (1962).

\bibitem{connect}
S.T.~Petcov and A.Yu.~Smirnov, Phys. Lett. B {\bf 322}, 109 (1994);\\
P.H.~Frampton, T.W.~Kephart, S.~Matsuzaki, Phys. Rev. D {\bf 78}, 073004 (2008);\\
Q.~Duret, B.~Machet, and M.I.~Vysotsky, Eur. Phys. J. C {\bf 61},
247 (2009);\\
W.~Rodejohann, Phys. Lett. B {\bf 671}, 267 (2009).



\bibitem{smirnov}
A.~Y.~Smirnov, arXiv:hep-ph/0402264;\\
H.~Minakata and A.Y.~Smirnov, Phys. Rev. D {\bf 70}, 073009 (2004).

\bibitem{raidal} M.~Raidal, Phys.\ Rev.\ Lett.\  {\bf 93},  161801
(2004).

\bibitem{phenomenology}
See, e.g.,


  P.~H.~Frampton and R.~N.~Mohapatra,
  JHEP {\bf 0501},  025 (2005)
  [arXiv:hep-ph/0407139];
  N.~Li and B.-Q.~Ma,
  Phys.\ Rev.\  D {\bf 71}, 097301 (2005)
  [arXiv:hep-ph/0501226];
  S.~Antusch, S.~F.~King and R.~N.~Mohapatra,
  Phys.\ Lett.\  B {\bf 618},  150 (2005)
  [arXiv:hep-ph/0504007];
  H.~Minakata,
  arXiv:hep-ph/0505262;
  J.~Ferrandis and S.~Pakvasa, Phys. Rev. {\bf D71}, 033004 (2005);
S.K.~Kang, C.S.~Kim and J.~Lee, Phys. Lett. {\bf B619}, 129 (2005);
S.~Antusch, S.F.~King and R.N.~Mohapatra, Phys. Lett. {\bf B618},
150 (2005); M.A.~Schmidt and A.Y.~Smirnov, Phys. Rev. {\bf D74},
113003 (2006);
  K.~A.~Hochmuth and W.~Rodejohann,
  Phys.\ Rev.\  D {\bf 75}, 073001 (2007)
  [arXiv:hep-ph/0607103];
F.~Plentinger, G.~Seidl and W.~Winter, Phys. Rev. {\bf D76}, 113003
(2007);
  G.~Altarelli, F.~Feruglio and L.~Merlo,
  JHEP {\bf 0905}, 020 (2009).


\bibitem{CK}
L.L.~Chau and W.Y.~Keung,
Phys. Rev. Lett. {\bf 53}, 1802 (1984).

\bibitem{9/12}
H. Fritzsch and Z.-Z. Xing,
  Phys.\ Rev.\  D {\bf 57}, 594 (1998).

\bibitem{Jarlskog}
C. Jarlskog, Phys.\ Lett.\ B {\bf 625}, 63 (2005).

\bibitem{Zheng10}
Y.-j.~Zheng, Phys.\ Rev.\ D {\bf 81}, 073009 (2010) [arXiv:1002.0919
[hep-ph]].


\bibitem{expt}
T2K Collaboration, K. Abe {\it et al.}, Phys. Rev. Lett. {\bf 107}, 041801 (2011);\\
MINOS Collaboration, P. Adamson {\it et al.}, Phys. Rev. Lett. {\bf 107}, 181802 (2011);\\ 
  Y.~Abe {\it et al.}  [DOUBLE-CHOOZ Collaboration],
  Phys.\ Rev.\ Lett.\  {\bf 108}, 131801 (2012)
  [arXiv:1112.6353 [hep-ex]].


\bibitem{pdg}
K. Nakamura {\it et al.} (Particle Data Group), J. Phys. G {\bf 37},
075021 (2010).

\bibitem{repara}
G.-N. Li, H.-H. Lin, X.-G. He, Phys. Lett. {\bf B 711}, 57 (2012) [arXiv:1112.2371 [hep-ph]].

\bibitem{global}
G. Fogli, E.~Lisi, A. Marrone, A.~Palazzo and A. M. Rotunno, Phys. Rev. D {\bf 84}, 053007 (2011).

\bibitem{Daya-Bay}
  F.P.~An {\it et al.}  [DAYA-BAY Collaboration],
  Phys.\ Rev.\ Lett.\  {\bf 108}, 171803 (2012).

\bibitem{Ahn:2012nd}
  J.K.~Ahn {\it et al.}  [RENO Collaboration],
 Phys.\ Rev.\ Lett.\ {\bf 108}, 191802 (2012). 


\bibitem{Zhang:2012xu}
  X.~Zhang and B.-Q.~Ma,
Phys. Lett. B {\bf 710},  630 (2012) [arXiv:1202.4258 [hep-ph]].

\bibitem{Zheng:2011uz}
  Y.-j.~Zheng and B.~-Q.~Ma,
  Eur.\ Phys.\ J.\ Plus {\bf 127}, 7 (2012)
  [arXiv:1106.4040 [hep-ph]].


\end{thebibliography}
\end{document}